\documentclass[useAMS,usenatbib]{mn2e}
\usepackage{lscape}
\usepackage{rotating}
\usepackage{amssymb}
\usepackage{float}
\usepackage{txfonts}
\usepackage{natbib}
\usepackage{graphicx}

\usepackage{tipa}
\usepackage{mathrsfs}
\usepackage{natbib,twoopt}

\def\mch{M$^{\rm c}$Hardy}
\def\rxte{{\it RXTE}}
\def\xmm{{\it XMM-Newton}}
\def\ergs{erg s$^{-1}$ cm$^{-2}$}

\def\cm3{\hbox{cm$^{-3}$}}

\input epsf.sty
\usepackage{graphicx}
\usepackage{epsfig}
\title[NGC 4051 time-lags]
{The long term X--ray time-lags of  NGC 4051}

\author[Papadakis et~al.]{I. E. Papadakis$^{1,2}$\thanks{E-mail:
jhep@physics.uoc.gr}, A. Rigas$^{3}$, A.  Markowitz$^{4,5},$ I. M. \mch$^{6}$ 
\\
\\
$^{1}$ Department of Physics and Institute of Theoretical and Computational Physics, University of Crete, 71003 Heraklion, Greece \\
$^{2}$ Foundation for Research and Technology - Hellas, IESL, Voutes, 71110 Heraklion, Greece \\
$^{3}$ Department of Materials Science and Technology, University of Crete, 71003 Heraklion, Greece \\
$^{4}$ Nicolaus Copernicus Astronomical Center, Polish Academy of Sciences, Bartycka 18, PL-00-716 Warszawa, Poland \\
$^{5}$ University of California, San Diego, Center for Astrophysics and Space Sciences, 9500 Gilman Dr., La Jolla, CA 92093-0424, USA \\
$^6$ Department of Physics and Astronomy, The University of Southampton, Southampton SO17 1BJ, UK
}
\date{Accepted. Received; in original form}
\pagerange{\pageref{firstpage}--\pageref{lastpage}}

\begin{document}
\maketitle
\begin{abstract}
We present the long term, frequency-dependent, X-ray time lags of the Seyfert galaxy NGC 4051.  We used {\it Rossi X-ray Timing Explorer (\rxte)} light curves in the 2--4, 4--7 and 7--10 keV bands and we measured the time-lags at $10^{-7}-10^{-6}$ Hz. This is the lowest frequency range that AGN time-lags have ever been measured. The variations in the higher energy bands are delayed with respect to the variations we observe at 2--4 keV, in agreement with the time-lags at high frequencies. When we combine our results with those from the model fitting of the time lags at higher frequencies we find that that the X--ray hard lags in NGC 4051 follow a power law of slope $\sim -1$ over a broad frequency range, from $\sim 10^{-7}$ to $\sim 10^{-3}$ Hz. This is consistent with the time-lags of Cyg X--1, a result which further supports the analogy between active galaxies and Galactic X--ray black hole binaries. 
\end{abstract}

\section{Introduction}
\label{sec:intro}

One of the defining characteristics of Active Galactic Nuclei (AGN) is that their flux is variable at all wavelengths. In general, the amplitude of the variations increase and the variability time scales decrease with increasing emission frequency. For example, the X--ray flux emitted from nearby Seyfert galaxies can vary by a factor of a few on time scales as small as a few hours, as opposed to time-scales of weeks/months and years at optical wavelengths. 

In principle, the study of the correlation between the variations observed in different energy bands can provide important clues regarding the variability mechanism and the source geometry. This study can be done with the use of the so-called correlation functions but, in the case of the X--ray variability studies of accreting objects, this is is usually performed in the Fourier space with the estimation of time-lags as a function of Fourier frequency. 

Such studies were first performed in black hole X--ray binaries (e.g. Miyamoto \& Kitamoto 1989; Nowak \& Vaughan 1996; Nowak et al. 1999; Pottschmidt et al. 2000). Time-lags were later reported in a few AGN as well (e.g. Papadakis, Nandra \& Kazanas 2001; \mch\ et al. 2004, 2007; Ar{\'e}valo et al. 2006;  Markowitz et al. 2007, Ar{\'e}valo, M$^{\rm c}$Hardy \& Summons 2008; Sriram, Agrawal \& Rao 2009). In all cases, the `hard'  band variations are delayed with respect to the variations at `softer' bands\footnote{In the case of the \rxte\ data, the $\sim 5-20$ keV band is usually considered as `hard', while `soft' is the $\sim 2-5$ keV band.}. These are the so-called `hard' or 'continuum' time-lags. `Soft' (or `reverberation') time-lags have also detected at frequencies higher than $\sim 5\times 10^{-4}$ Hz.  In this case, it is the soft band photons that are delayed with respect to the hard band photons. They are probably caused by light travel time delays between variations in the primary X--ray emission and the corresponding variations in the X--ray reflection emission from the inner accretion disc. They are usually detected between energy bands below $\sim 1$ keV or around $5-7$ keV (where the contribution of the X--ray reflection component is expected to be higher) and $\sim 2-4$ keV (which is representative of the primary radiation).

Recently, Epitropakis \& Papadakis (2017; EP17 hereafter) studied in detail the X--ray continuum time-lags (and intrinsic coherence) between 2--4 keV and various other energy bands in the 0.3--10 keV range for 10 X-ray bright and highly variable AGN. They used \xmm\, data, and they estimated the time-lags in the frequency range between $5\times 10^{-5}$ Hz and $\sim 3-6\times 10^{-4}$. They found that higher energy bands are always delayed with respect to the low energy ones, and the time-lags show a power-law dependence on frequency with a slope of $-1$. Their amplitude scales with the logarithm of the energy separation between the light curves, and it increases with the square root of the ratio of the X--ray over the Eddington luminosity. 

Like EP17, most of the past studies of the continuum time--lags in AGN used data from one or a few \xmm\, orbits. NGC 7469 and Ark564 are exceptions, since Papadakis et al. (2001) used $\sim 1$ month long long \rxte\,  light curves and Ar{\'e}valo et al. (2006) used similarly long {\it ASCA} data, respectively. In this work, we use much longer \rxte\, light curves in order to study the continuum time-lags of NGC 4051 at very low frequencies.  We chose NGC 4051 (which is a Seyfert 1 galaxy) because it displays rapid and high-amplitude X-ray flux and spectral variability (see e.g. Lawrence et al. 1987; Papadakis \& Lawrence 1995;  Uttley et al. 1999; \mch\ et al. 2004; Ponti et al. 2006; Terashima et al. 2009; Vaughan et al. 2011). In addition, it was observed many times by \rxte\, and the resulting light curves are the longest and best-sampled light curves among all AGN monitored by \rxte. Part of these light curves were used by \mch\, et al. (2004) to estimate the power spectral density function (PSD) over  the broadest frequency range for any AGN. PSD and time-lags measurements at such low frequencies can, in principle, put interesting constrains in theoretical models. For example, in the context of the propagating accretion rate fluctuations model (Lyubarskii 1997), the low-frequency variability amplitude and time-lags can be used to identify the radius at which the longest X--ray variations are generated. 

\mch\, et al.\, (2004) were the first to study the X--ray time-lags in NGC 4051. They used data from a single \xmm\, orbit (in 2001) and they detected hard lags at frequencies $1.3\times 10^{-5}- 5\times 10^{-4}$ Hz. Alston et al (2013) used multiple \xmm\, observations taken in 2009, while De Marco et al (2013)  used both the 2001 and 2009 \xmm\, data to study the X-ray time-lags of the source. Both papers focused on the study of the soft, reverberation time lags at high frequencies (i.e. above $10^{-4}$ Hz) using light curves in the $0.3-1$ keV and $2-5$ keV bands. In this work we use the \rxte\, light curves of NGC 4051 and the advanced techniques of Epitropakis \& Papadakis (2016; EP16 hereafter) to estimate the time-lags at frequencies between $\sim 10^{-7}$ and $10^{-6}$ Hz. This is the lowest frequency that time-lags in AGN have ever been measured. 

In the following section we discuss briefly the data analysis. In the third section we describe the method we used to estimate the time lags and we present our results.  In the last section we compare the low frequency time-lags with the time-lags estimates at higher frequencies, and we discuss some implications of our results. 

\section{Data analysis}  \label{sec:sample}

NGC 4051 is the most observed AGN with \rxte, with more than 2100 observations. The observations span over a period of approximately 16 years. The first observation took place on 1996-04-23 and the last on 2011-12-30. The exposure time for most of them is $\sim 1-3$ ks. We used 2--4, 4--7 and 7--10 keV light curves from the "\rxte\ AGN Timing \& Spectral Database" (https://cass.ucsd.edu/$~$rxteagn/).  We discuss briefly the basic data analysis below. Full details can be found at the webpage of this Database as well as in Rivers, Markowitz \& Rothschild (2011, 2013).

STANDARD-2 data from the Proportional Counter Array (PCA) were used. In particular, Proportional Counter Units (PCUs) 0, 1 and 2 were used prior to 1998 December 23. Data from PCUs 0 and 2 were used from 1998 December 23 until 2000 May 12, and data from PCU 2 only were used after 2000 May 12. Events from the top Xe layer only were selected in order to maximize signal-to-noise. Standard screening was applied: data were rejected if taken within 20 minutes of the spacecraft's passing through the South Atlantic Anomaly, if {\tt ELECTRON0 > 0.1} ({\tt ELECTRON2} for data after 2000 May 12), if the spacecraft was pointed within 10 degrees of the Earth, or if the source was $> 0.02$ degrees from the optical axis. The 2--10 keV PCA spectrum was extracted for each observation using HEASOFT version 6.7 and the ``rex" perl script. The  (background subtracted) observed spectrum was then fitted in the 2--4, 4--7 and 7--10 keV sub-bands with a power-law model, and  the source's flux was estimated in each one of these bands. 

\begin{figure*}
\centering
 \includegraphics[bb=71 15 550 770,width=9.6cm,angle=270,clip]{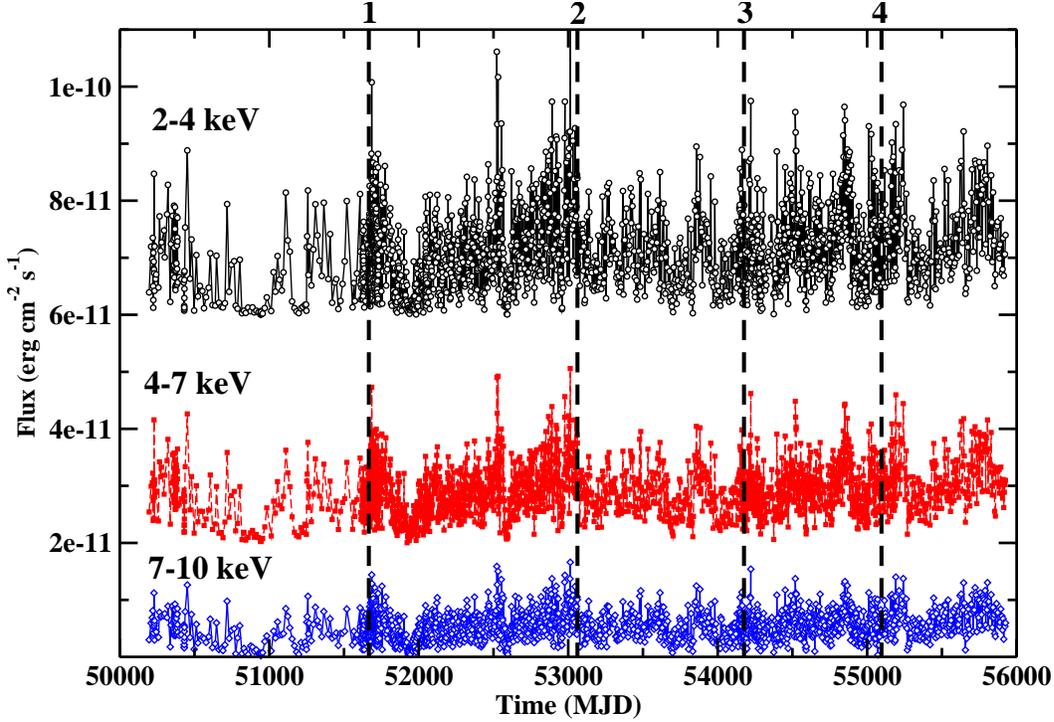}
\caption{The \rxte, 2--4, 4--7 and 7--10 keV band light curves of NGC 4051 (before binning them). For clarity reasons, we have shifted the 4--7 and 2--4 keV band light curves by adding 2 and $5\times 10^{-11}$ \ergs, rspectively. The vertical dashed lines indicate the light curve parts that we binned using a bin size of 2 days (see text for details). }
\label{fig:lcs}
\end{figure*}

Figure \ref{fig:lcs} shows the \rxte\ 2--4, 4--7 and 7--10 keV band light curves (black open circles, red filled squares and open blue diamonds, respectively). The error on each flux point was obtained by dividing the standard deviation of the $N$, 16-s binned count rate light curve points in each observation by $\sqrt N$. At the start of the monitoring program, and for approximately 1,5 years, \rxte\ observed the source every 8-11 days (on average). After MJD $51626$ (June 2000) and until the end of the mission, the observations became more frequent. In some periods, observations were taken once or even twice per day. For most of the time,  observations were taken once every 2 - 4 days. 

Clearly, the source is highly variable on all sampled time scales. Furthermore, similar variations appear in all the three bands. These light curves are ideal in order to estimate time-lags between the three energy bands, on very long time scales. 

\section{Time-lag estimation}  \label{sec:lags}

We estimated the 2--4 vs 4--7 keV and the 2--4 vs 7--10 keV time-lags (the ``S-vs-H1" and ``S-vs-H2" time-lags, respectively) following EP16. Their techniques assume evenly sampled light curves. Although \rxte\ observed NGC 4051 quite frequently, the resulting light curves are not evenly sampled. For that reason, we first binned various parts of the light curves using a bin size of $\Delta t=2$ and $\Delta t=4$ days, depending on the average separation between successive observations in these parts. We used a bin size of 2 days in the time intervals between MJD $51666.4-53061.6$ (start/end dates are indicated by the vertical dashed lines1 and 2 in Fig.\,\ref{fig:lcs}) and MJD $54176.6-55095.1$ (lines 3 and 4 in Fig.\,\ref{fig:lcs}). We used a  bin size of 4-days for the data in the period between MJD $53137.0-54141.4$ (this is the period between the vertical lines 2 and 3 in Fig.\,\ref{fig:lcs}), and towards the end of the observations (MJD $55201.5 - 55935.5$). The duration of the 2 days binned light curves is 2314 days, while the duration of the 4 days binned light curve parts is 1738 days. Their total duration is 4052 days, which is $\sim 71$\% of the duration of the original data set.

The flux and error in each bin are set to be equal to the straight mean of the flux and error of all the observations within the bin.  The time stamp of the binned light curves, $t_{bin}$, was set to the center of each bin. On average, the difference between $t_{bin}$ and the mean of the start time of all observations within each bin was less than 0.3 days. This is small compared to the bin size, and much smaller than the time scales over which we estimate time-lags. There are a few missing points in the binned light curves. We added points using linear interpolation to fill the gaps, as long as there were less than 3 consecutive point missing. We randomized the inserted fluxes, by adding Gaussian noise with a standard deviation equal to the average error of all the points in the binned light curves. 

Following EP16, reliable estimation of the time-lags requires the use of a large number of light curves segments. To this end, we divided both the 2 and the 4 days binned light curves into segments of duration $T=100$ days. There are $m_{2days-binned}=19$ and $m_{4days-binned}=18$ such segments.

We used equations (9) and (10) in EP16 to calculate the cross-periodogram of all segments in the 2--4/4--7 keV and 2--4/7-1-10 keV bands, at $\nu_p=p/(N\Delta t)$, where $p=1,...,N/2$, $N$ being the number of points in each segment. We adopted, 
\noindent
\begin{equation} \label{eq:eq1}
\hat{C}_{xy}(\nu_p)=\frac{1}{m}\sum_{k=1}^{m}I^{(k)}_{xy}(\nu_p),
\end{equation}
\noindent
and
\noindent
\begin{equation} \label{eq:eq2}
\hat{\tau}_{xy}(\nu_p)\equiv\frac{1}{2\pi\nu_p}\mathrm{arg}[\hat{C}_{xy}(\nu_p)],
\end{equation}
\noindent
as our estimates of the cross-spectrum and of the time-lag spectrum, respectively. We adopted the standard convention of defining $\mathrm{arg}[\hat{C}_{xy}(\nu_p)]$ on the interval $(-\pi,\pi]$. $I^{(k)}_{xy}(\nu_p)$ in eq.\,(1) is the cross-periodogram of the $k-$th segment at frequency $\nu_p$, and the letter $m$ denotes the number of segments. To increase the statistics and reduce the error on the time-lags estimates, we combined the 2 and 4-days binned segments together so that $m_{all}=m_{2days-binned}+m_{4days-binned}=37$. Since the number of points is different in the 2 and 4-days binned segments, the highest frequency at which we estimated the combined time-lag spectrum is the Nyquist frequency for the 4-days binned segments, which is equal to $1/(2\times 4 {\rm days})=1.45\times 10^{-6}$ Hz. 

 The analytic error estimate of $\hat{\tau}_{xy}(\nu_p)$ is given by, 
 \noindent
\begin{equation} \label{eq:eq3}
\sigma_{\hat{\tau}}(\nu_p)\equiv\frac{1}{2\pi\nu_p}\frac{1}{\sqrt{2m}}\sqrt{\frac{1-\hat{\gamma}^2_{xy}(\nu_p)}{\hat{\gamma}^2_{xy}(\nu_p)}},
\end{equation}
\noindent
where, 
\noindent
\begin{equation} \label{eq:eq4}
\hat{\gamma}^2_{xy}(\nu_p)\equiv\frac{|\hat{C}_{xy}(\nu_p)|^2}{\hat{P}_x(\nu_p)\hat{P}_y(\nu_p)}.
\end{equation}
\noindent
Equation\,\ref{eq:eq4} defines the estimator of the so-called coherence function. $\hat{P}_x(\nu_p)$ and $\hat{P}_y(\nu_p)$ are the periodograms of the light curves, which are also calculated by binning over $m$ segments. 

\begin{figure}
\centering
 \includegraphics[bb=70 5 550 770,width=5.5cm,angle=270,clip]{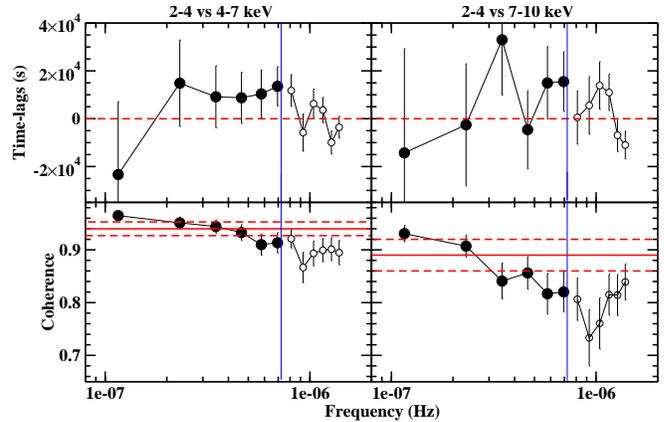}
\caption{Top panels: The 2--4\,keV vs. 4--7\,keV (S-vs-H1) and the 2--4\,keV vs 7--10\,keV (S-vs-H2) time-lags vs frequency plots (left and right, respectively). Bottom panels: The respective plots of the coherence functions. The blue, continuous vertical lines (in all panels) indicate the highest frequency up to which we can reliable estimate the time-lags. The horizontal solid and dashed lines in the bottom panels indicate the EP17 results (see text for details).}
\label{fig:timelags}
\end{figure}

The resulting S-vs-H1 and S-vs-H2 time-lags are plotted on the left and right top panels of Fig.\,\ref{fig:timelags}, respectively. Positive time-lags indicate that the higher energy band variations follow those in the 0.2--4 keV band. The respective coherence functions are plotted in the bottom panels of the same figure.  There is a maximum frequency, $\nu_{\rm{max}}$, at which we can reliably estimate time-lags (EP16). This frequency depends on the amplitude of the sample coherence, $\hat{\gamma}^2_{xy}(\nu_p)$. Even if the intrinsic coherence is equal to one at all frequencies, the sample coherence will decrease exponentially towards zero at high frequencies, due to the Poisson noise in the light curves. EP16 suggest to estimate the time-lags up to frequency where $\hat{\gamma}^2_{xy}(\nu_p)=1.2/(1+0.2m)$. In our case, $1.2/(1+0.2m)=0.14$. As the bottom panels in Fig.\,\ref{fig:timelags} show, the sample coherence functions in our case are significantly higher than this number, at all frequencies. In this case, $\nu_{\rm{max}}$ is mainly determined by the aliasing effects. 

Aliasing in the case of the cross-spectrum estimation has the effect of decreasing the imaginary and increasing the real part of the cross-periodogram. As a result, their ratio (which determines time-lags) is biased.  Due to this effect, time-lags should be estimated at frequencies lower than 1/5 of the Nyquist frequency, in the case of a sampled light curve, or half the Nyquist frequency for binned light curves. In our case, the light curve segments that we used are binned, although the number of the individual \rxte\ observations in each bin is relatively small (of the order of $\sim 2-6$, typically). The vertical, continuous blue lines in Fig.\,\ref{fig:timelags} indicate the frequency which is equal to $0.5\times \nu_{\rm Nyquist}$ of the 4-days binned segments. This is 4 times smaller than the Nyquist frequency of the 2-days binned segments (thus approaching the value that EP16 suggests, even in the case of the sampled light curves).  We accept this as the highest frequency at which we can reliably estimate time-lags. 

The S-vs-H1 time-lags below $\nu_{\rm max}$ are all positive (except the lowest frequency estimate, which has a large error). Their weighted mean is $10200\pm 4700$ s. The S-vs-H2 time-lags have a larger error, because the amplitude of the S-vs-H2 coherence is smaller than the amplitude of the S-vs-H1 coherence (see the bottom panels in Fig.\,\ref{fig:timelags}). The weighted mean of the S-vs-H1 time-lags below $\nu_{\rm max}$ is 9910$\pm 7070$ s.

\section{Summary and discussion} \label{sec:discussion}

We have used long-term \rxte\ flux light curves in three energy bands, and the techniques of EP16, to estimate the continuum, low frequency, X--ray time-lags in NGC 4051. These are the longest and best sampled X--ray light curves among all AGN monitored by \rxte. Given their duration, sampling rate, as well as the fact that NGC 4051 is among the most variable Seyferts in X--rays,  these light curves are ideal for the estimation of the time-lags at low frequencies.  
We managed to reliably estimate the time-lags over almost a decade in frequency, from $\sim 10^{-7}$ up to $\sim 7\times 10^{-7}$ Hz. This is the lowest frequency range at which X--ray time-lags have ever been measured in AGN. The time-lags we detected imply that the 4--7 and 7--10 keV band variations are delayed with respect to the variations seen in the 2--4 keV band. This in agreement with the results from previous studies which also found hard continuum lags, but at higher frequencies.


\begin{figure}
\centering
 \includegraphics[bb=70 0 560 690,width=6.4cm,angle=270,clip]{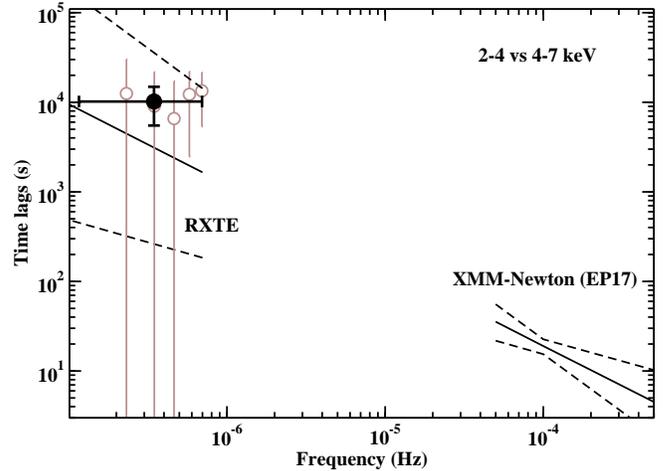}
\caption{The continuous solid line indicates the EP17 best-fit to the S-vs-H1 time-lags spectrum (at high frequencies), and its extrapolation to the frequency range where we estimate the \rxte\ time-lags. The dashed lines show the $1\sigma$ uncertainty, using the combined $1\sigma$ error on the slope and normalisation. Empty circles show the \rxte\ time-lags, and the filled circle shows their mean (plotted at the mean of the logarithm of frequencies).}
\label{fig:results}
\end{figure}

\mch\ et al. (2004) were the first to estimate X--ray, continuum time-lags in NGC 4051. They used the 2001 \xmm\, data and they found that hard photons were lagging the soft photons. The time-lags spectra were consistent with a power law with a slope of $-1$, and their amplitude increased as the energy separation between bands increased. EP17 also studied the continuum time-lags in NGC 4051 using all the archival \xmm\, data of the source. They estimated time-lags between various energy bands and  2--4 keV, and they fitted them with a power-law model in the frequency range from $5\times 10^{-5}$ Hz to $5\times 10^{-4}$ Hz. They found that the X--ray continuum time-lags in this source can be fitted well with a function of the form: $\tau(\nu, E, {\rm 3 keV})=73(\pm 14)\log(E/{\rm 3 keV}) (\nu/10^{-4})^{-0.9\pm0.4}$ sec, where $E$ is the mean of the energy band that one wishes to correlate with 2--4 keV (the mean of this band was set to 3 keV).

The continuous, solid line at high frequencies in Fig.\,\ref{fig:results} shows the EP17 best-fit to the 2--4 vs 4--7 time-lags at high frequencies. It also shows its extrapolation to low frequencies. The empty circles show the S-vs-H1 time lags, and the filled circle shows their mean. Our results are consistent with an extrapolation of the time-lag spectra derived at higher frequencies down to below $10^{-6}$. We conclude that the X--ray continuum time-lags in NGC 4051 are consistent with a power law of slope $\sim -1$ over a broad frequency range, from $\sim 10^{-7}$ to $\sim 5\times 10^{-4}$ Hz. The S-vs-H2 time-lags (plotted in the upper right panel of Fig.\,\ref{fig:results}) are also consistent with the extrapolation of the EP17 best-fit line to the 2--4 vs 7--10 keV time-lags, but we do not show a similar plot in this case because of the larger error of the S-vs-H1 time-lags (which is due to the lower coherence in this case). 

Although the average time lags in NGC 4051 (i.e. the time lags estimated using all the \xmm\ data) show clear hard lags at low frequencies (EP17),  there is also a suggestion that they may be flux dependent (Alston et al 2013). These authors studied the time lags between 0.3--1 and 2--5 keV bands. The flux dependence of the time-lags between the same bands that we use in this work can be seen in Alston PhD thesis (2013).  The leftmost panels in their Figure 6\,.3 shows that, at low frequencies ($0.5-1.5\times 10^{-4}$ Hz), the 4--7 and 7--10 keV bands lag behind the 2--4 keV band at  high fluxes. The situation appears to reverse at medium and low fluxes, where the  2-4 keV band tends to lag behind the 4-7 and 7-10 keV bands. 

The X--ray reverberation signal should not contribute significantly to the observed time lags at such low frequencies, although the warm absorber may contribute to the ``dissappearence"  of the hard lags at low frequencies when the source's flux decreases.  Silva, Uttley \& Constantini (2016) have shown that warm absorbers can introduce (negative) time lags between the more absorbed bands relative to the continuum caused by  the  response  of  the  gas  to  the continuum variations. These time lags are more pronounced during the medium and low-flux states, as observed. Such an effect could easily account for the flux-related time lags changes between the softest bands (below 1 keV) and the continuum but it  may not be able to explain the Alston (2013) results at high energies in full. In any case,  the available  \rxte\ data are not sufficient to perform a flux resolved analysis at low frequencies. The error of the observed time lags is quite large even when we use all the available data (see below), so any attempt to investigate flux-related time lag variations will not be meaningful.

In general, the errors of the S-vs-H1 and S-vs-H2 time-lags are large, despite the fact that the \rxte\ light curves resulted in 38, 100 days long segments. This is larger than the number of segments that EP17 used. According to EP16, as long as the number of segments is larger than 20,  the time-lags estimates should be approximately Gaussian, and equation 3 should provide a reliable estimate of their error. This equation shows that the error decreases with the square root of the number of segments. Therefore, in order to decrease the error, say by a factor of two, we will need to increase the number of segments by a factor of four. However, since it is highly unlikely that there will be any light curves longer than the NGC 4051 \rxte\ light curves in the next years, it will not be possible to reduce the error of the time-lags at at low frequencies (unless we find new techniques to estimate them from the same data sets). 

The second parameter that determines the time-lag error is the value of the coherence. The closer to one, the smaller the time-lags error will be. We found that the coherence between 2--4 and 4--7 keV light curves is rather flat, with a value of $\sim 0.94$ at all frequencies. Similarly, the 2--4 vs 7--10 keV coherence is $\sim 0.9$. We observe a slight decrease of the coherence with increasing frequency, probably due to Poisson noise (the decrease of coherence with increasing frequency is more pronounce in the  S-vs-H2 case). EP17 also found that the intrinsic coherence is constant at high frequencies, and that it decreases with increasing energy separation. The horizontal solid lines in the bottom panels of Fig.\,\ref{fig:timelags} show the EP17 best fit coherence (dashed lines show the $\pm 1\sigma$ errors). Our estimates are fully consistent with the high-frequency coherence measurements, and indeed support the hypothesis that the coherence decreases with increasing energy separation in this source.


\mch\ et al. (2004) found that the overall long and short time-scale power-spectral density function (PSD) of NGC 4051 is a close match to the PSD of Cygnus X-1 in the high/soft state. The main argument was that the PSD slope below the high-frequency break remains unchanged for over four decades.  In this work, we found strong indications that the time-lags spectrum of NGC 4051 also remains unchanged over 3.5-4 decades in frequency.  This is very similar to the time-lags spectrum of Cyg X--1 in its high/soft state (Pottschmidt et al. 2000). Indeed the EP17 best-fit model (which as we argued above agrees very well with the low-frequency time-lags spectrum of NGC 4051) predicts a time-lag of $\sim 0.01$ s at 1 Hz between 2--4 and 8--13 keV. This is almost exactly what we observe in the time-lags spectrum of Cyg X--1 (see e.g. right panel in Fig.\,4 of Pottschmidt et al., 2000).  

The X--ray time--lag spectra of NGC 4051 and NGC 7469 are very similar. They both show a power-law shape with a slope of $\sim -1$ down to very low frequencies. The PSDs of the two sources are also similar. A power-law of slope $\sim -1$ at low frequencies, which becomes steeper at frequencies above a bending frequency can fit them well (\mch\ et al. 2004, Markowitz 2010). These similarities suggest that both sources operate in the same state (most probably in the high/soft state, given the similarity of their time-lags spectra and PSDs with those of Cyg X-1). On the other hand, the NGC 4051 time-lag spectrum is different than the Ark564 time-lag spectrum. The same is true for their PSDs. The Ark564 PSD shows two bending frequencies, and it can be fitted well by the sum of two Lorentzians (\mch\, et al. 2007), contrary to NGC 4051, which does not show any low frequency bending down to $10^{-7}$ Hz.  Similarly, the continuum time-lags in Ark564 are approximately constant in the frequency range below $10^{-5}$ Hz,  which is not the case with NGC 4051. These differences imply that the two sources operate at different states. This is probably due to the accretion rate being significantly different in these sources (close to the Eddington limit in Ark564, significantly smaller in NGC 4051). 

Both the PSD and the time-lags spectra of Cyg X--1 in the high/soft state can be explained by propagating mass accretion rate fluctuations model (Lyubarskii 1997; Kotov, Churazov \& Gilfanov 2001). According to this model, the X-ray variations over a broad range of time scales arise because of mass accretion rate fluctuations stirred up at each radius of the accretion flow and propagating towards the black hole. If the spectrum of the region close to the BH is harder than that emitted at larger radii, the propagation of fluctuations also produces hard time-lags.  Ar{\'e}valo \& Uttley (2006) and Rapisarda, Ingram \& van der Klis (2017) have actually fitted the model to the PSDs and time-lags spectra of Cyg X--1 in soft state. They found that the data can be explained in the case when a stable disc is sandwiched above and below by a variable hot flow, which emits the X--rays. Accretion rate fluctuations are stirred up and propagating only in the hot flow, which has a very large radial extent (i.e. up to 2500 gravitational radii, according to the Rapisarda et al. best fit results). Although the emissivity profile of this hot flow is very steep (so that most of the X--rays are emitted close to the BH), the large radial extend is needed in order to explain the large amplitude variability and long delays at low frequencies. It is difficult to understand the physical mechanism responsible for the creation of this hot flow (with temperatures up to $\sim 10^8-10^9$ K) at such large radii. In such a case, one has to also take into account the time-lags due to the repeated scattering of low-energy photons by the hot electrons in a very extended corona (albeit in the radial direction),  as they may be of sizeable amount. 

An alternative interpretation for the X--ray continuum time delays in AGN has been put forward by Miller et al (2010).  According to their model, hard band time lags at low frequencies could arise if part of the X-ray continuum is scattered from a uniform, isotropic, shell. In this case, the hard lags should remain constant at frequencies smaller than the frequency which corresponds to the maximum time delay (this is the light travel time across the diameter of the shell). Such a model has been fitted by Turner et al (2017) to the 2--4 vs 5--7.5 keV time lags of NGC 4051, estimated with {\it NuSTAR} data. According to their model fits, the time lags at  frequencies below $\sim 10^{-4}$ Hz  should be equal to $\sim 400-500$ s (bottom panel of their Fig.\,3). Strictly speaking the time lags below $\nu_{\rm max}$ in the top-left panel in Fig.\,\ref{fig:timelags} are consistent with this prediction. However, we notice that (almost) all time lags are positive and larger than this value. Using the binomial probability distribution, the probability of five consecutive time lags to be larger than the expected value is rather low (i.e. $\sim 3$\%), suggesting that this is not a likely explanation of the observed time lags.

\section*{Acknowledgments}

This work has made use of lightcurves provided by the University of California, San Diego Center for Astrophysics and Space Sciences, X-ray Group (R.E. Rothschild, A.G. Markowitz, E.S. Rivers, and B.A. McKim), obtained at http://cass.ucsd.edu/$~$rxteagn/.

\vspace{-0.5cm}
\section*{References}
Alston W.~N., 2013, PhD thesis, Univ. Leicester (https://lra.le.ac.uk/handle/2381/28951)\\
Alston W.~N., Vaughan S., Uttley P., 2013, MNRAS, 435, 1511 \\
Ar{\'e}valo P., Papadakis I.~E., Uttley, P., \mch\ I.~M., Brinkmann W., 2006, MNRAS, 372, 401 \\
Ar{\'e}valo P., Uttley P., 2006, MNRAS, 367, 801 \\
Ar{\'e}valo P., \mch\ I.~M., Summons D.~P., 2008, MNRAS, 388, 211 \\
De Marco, B., Ponti, G., Cappi, M., Dadina M., Uttley P., Cackett E.~M., Fabian A.~C., Miniutti G., 2013, MNRAS, 431, 2441 \\
Epitropakis A., Papadakis I.~E., 2016, A\&A, 591, 113 (EP16)\\
Epitropakis A., Papadakis I.~E., 2017, MNRAS, 468, 3568 (EP17)\\
Kotov O., Churazov E., Gilfanov M., 2001, MNRAS, 327, 799
Lawrence A., Watson M.~G., Pounds K. ~A.,  Elvis M., 1987, Nature, 325, 694 \\
Lyubarskii Y. E., 1997, MNRAS, 292, 679 \\
Markowitz A., Papadakis I., Ar{\'e}valo P., Turner T.~J., Miller L., Reeves J.~N., 2007, ApJ, 656, 116 \\
Markowitz A., 2010, ApJ, 724, 26 \\
Miller L., Turner T.~J.,  Reeves J.~N., Lobban A., Kraemer S.~B, Crenshaw D.~M., 2010, MNRAS, 403, 196 \\
Miyamoto S., Kitamoto S.,1989, Nature,  342, 773 \\
\mch\ I.~M., Papadakis I.~E., Uttley P., Page M.~J., Mason K.~O, 2004, MNRAS, 348, 783 \\
\mch\ I.~M.; Ar{\'e}valo P., Uttley P., Papadakis I.~E., Summons D.~P., Brinkmann W., Page M.~J., 2007, MNRAS, 382, 985  \\
Nowak M.~A., Vaughan B.~A., 1996, MNRASA, 280, 227 \\
Nowak M.~A., Vaughan B.~A., Wilms J., Dove J.~B., Begelman, M.~C., 1999, ApJ, 510, 874 \\
Papadakis I.~E., Lawrence A., 1995, MNRAS,  272, 161 \\
Papadakis I.~E., Nandra K., Kazanas D., 2001, ApJ, 554, 133 \\
Ponti G., Miniutti G., Cappi M., Maraschi L., Fabian A.~C., Iwasawa K., 2006, MNRAS, 368, 903 \\
Pottschmidt K., Wilms J., Nowak M.~A., Heindl W.~A., Smith D.~M., Staubert R., 2000, A\&A, 357, L17 \\
Rapisarda S., Ingram A., van der Klis M., 2017, MNRAS, 472, 3821 \\
Rivers E., Markowitz A., Rothschild R., 2011, ApJS, 193, 3 \\
Rivers E., Markowitz A., Rothschild R., 2013, ApJ, 772, 114 \\
Silva C.V., Uttley U., Constantini E., 2016, A\&A, 596, 79 \\
Sriram K., Agrawal V.~K., Rao A.~R., 2009, ApJ, 700, 1042 \\
Terashima Y et al., 2009, PASJ, 61, 299 \\
Turner T.J., Miller L., Reeves J. N., Braito V., 2017, MNRAS, 467, 3924 \\
Uttley P., McHardy I.~M., Papadakis I.~E., Guainazzi M., Fruscione A., 1999, MNRAS, 307, L6 \\
Vaughan S., Uttley P., Pounds K.~A., Nandra K., Strohmayer T.~E., 2011, MNRAS,  413, 2489 \\
 
\end{document}